# Automatic Inverse Treatment Planning for Gamma Knife Radiosurgery via Deep Reinforcement Learning

**Yingzi Liu[1], Chenyang Shen[2], Tonghe Wang[1], Jiahan Zhang[1], Xiaofeng Yang[1], Tian Liu[1], Shannon Kahn[1], Hui-Kuo Shu[1], Zhen Tian[1]**

[1]Department of Radiation Oncology, Emory University, Atlanta, GA 30022

[2]Department of Radiation Oncology, University of Texas Southwestern Medical Center, Dallas, TX 75235

Emails: Zhen.Tian@emory.edu

**Purpose**: Several inverse planning algorithms have been developed for Gamma Knife (GK) radiosurgery to determine a large number of plan parameters via solving an optimization problem, which typically consists of multiple objectives. The priorities among these objectives need to be repetitively adjusted to achieve a clinically good plan for each patient. This study aimed to achieve automatic and intelligent priority-tuning, by developing a deep reinforcement learning (DRL) based method to model the tuning behaviors of human planners.

**Methods:** We built a priority-tuning policy network using deep convolutional neural networks. Its input was a vector composed of the plan metrics that were used in our institution for GK plan evaluation. The network can determine which tuning action to take, based on the observed quality of the intermediate plan. We trained the network using an end-to-end DRL framework to approximate the optimal action-value function. A scoring function was designed to measure the plan quality.

**Results:** Vestibular schwannoma was chosen as the test bed in this study. The number of training, validation and testing cases were 5, 5, and 16, respectively. For these three datasets, the average plan scores with initial priorities were $3.63 \pm 1.34$, $3.83 \pm 0.86$ and $4.20 \pm 0.78$, respectively, while can be improved to $5.28 \pm 0.23$, $4.97 \pm 0.44$ and $5.22 \pm 0.26$ through manual priority tuning by human expert planners. Our network achieved competitive results with $5.42 \pm 0.11$, $5.10 \pm 0.42$, $5.28 \pm 0.20$, respectively.

**Conclusions:** Our network can generate GK plans of comparable or slightly higher quality comparing with the plans generated by human planners via manual priority tuning. The network can potentially be incorporated into the clinical workflow to improve GK planning efficiency.

## 1. Introduction

Gamma Knife (GK) radiosurgery, which delivers high conformal irradiation dose to selected intracranial targets while sparing the surrounding normal brain tissues, is an important and safe alternative to traditional neurosurgery for a variety of brain disorders, such as brain tumors, arteriovenous malformations, vestibular schwannomas, and meningiomas. Treatment planning is of vital importance to achieve the desired dose distribution to ensure treatment efficacy.

A GK treatment plan is composed of radiation shots, and one shot means to treat the target at an isocenter with a combination of gamma ray beams coming from eight sectors. Each sector has either 4-, 8-, or 16-mm available collimation size or complete beam blocking. During the treatment planning to sculpt the desired dose distribution, there is a high degree of freedom of choosing the number of shots, location of shot isocenter, sector collimation, and beam-on time for each shot. Using manual forward planning to place the shots and adjust those plan parameters is cumbersome and time-consuming. To ease the planning process and better exploit the high degree of freedom available in GK planning, multiple inverse planning algorithms have been developed for GK radiosurgery to determine the plan variables via mathematically solving an optimization problem [1-10]. The optimization problem of inverse planning typically consists of multiple objective terms designed for various clinical or practical considerations, such as target coverage, selectivity, gradient index, dose to organ at risks (OAR), and total beam-on time. Priorities among these planning objectives critically affect the resulting plan quality. Although optimization engines can solve the optimization problem for a given set of priority values, due to individual anatomy variations, planners still need to repeatedly interact with the optimization solver and adjust the objective priorities for each specific patient during GK inverse planning to make it clinically optimal, which is similar to the inverse planning for linac-based radiotherapy. This priority tuning process is usually a trial-and-error process and is time-consuming. In addition, the resulting plan quality is highly subjective to planners' experience and can be affected by the available planning time. GK radiosurgery is typically a one-day procedure, during which a stereotactic frame is usually needed to be attached to patients' skull to serve as a reference coordinate for treatment planning and a stabilization device for treatment delivery. This places a heavy demand on the planner to generate an optimal plan in a timely manner. Hence, there is a strong desire to develop automatic approaches to determine the priorities for inverse planning of GK radiosurgery.

Over the years, significant efforts have been devoted to automate the priority tuning process for linac-based radiotherapy. A commonly used approach is to add an outer loop of priority optimization on top of the treatment plan optimization problem with a fixed set of priorities, and adjust the priorities in the outer loop using greedy methods [11-14], heuristic methods [15,16], fuzzy inference [17-19], and statistical methods [20-22]. Recently, reinforcement learning (RL) has been brought into the radiotherapy field to solve the automatic priority tuning problem from a new angle, that is, by establishing an action-value function via an end-to-end RL process to model the intelligent adjustment behaviors of human expert planners during inverse planning. This function takes the state of the intermediate plan as input, and outputs a tuning action that will maximize the reward, which is usually correlated with plan quality improvement. The form of the optimal action-value function is usually unknown. A linear function has been used to approximate the optimal action-value function for automatic planning of stereotactic body radiation therapy for pancreas cancer [23]. However, this linear approximation may potentially limit the flexibility of the model when trying to approximate the underlying action-value function for complicated planning tasks. Deep neural network, on the other hand, possesses high flexibility and capacity to approximate complicate functions. By employing deep neutral networks to parameterize the optimal action-value function and trained the network via RL, Shen *et al.* has successfully built deep reinforcement learning (DRL) based virtual planners for high-dose-rate (HDR) brachytherapy[24] and intensity modulated radiotherapy (IMRT)[25-27] planning,

Inspired by these successes, in this study we explored the feasibility of developing a DRL-based virtual planner for GK radiosurgery. We focus on vestibular schwannoma cases in this study, as they are usually considered to be the most challenging for GK planning due to the irregularly-shaped target and its proximity to brainstem and cochlea.

## 2. Methods

### 2.1 GK treatment plan optimization model

In this study, the linear programming model proposed by Sjölund *et al.*[8] was used as the treatment plan optimization model in the DRL framework, since a recent study reported that linear programming tended to yield shorter beam-on time than convex quadratic penalty approaches and convex moment-based approaches [10]. The objective function for vestibular schwannoma cases is formulated as

$$
\begin{aligned}
\text{minimize}_t \quad & \frac{\omega_{TH}}{N_T} \sum_{i \in I_T} \max(D_i - D_{TH}, 0) \quad + \frac{\omega_{TL}}{N_T} \sum_{i \in I_T} \max(D_{TL} - D_i, 0) \\
& + \frac{\omega_{IS}}{N_{IS}} \sum_{i \in I_{IS}} \max(D_i - D_{IS}, 0) \quad + \frac{\omega_{OS}}{N_{OS}} \sum_{i \in I_{OS}} \max(D_i - D_{OS}, 0) \\
& + \frac{\omega_{BS}}{N_{BS}} \sum_{i \in I_{BS}} \max(D_i - D_{BS}, 0) \\
& + \omega_{CO} \max\left( \left( \sum_{i \in I_{CO}} \frac{D_i}{N_{CO}} \right) - D_{CO}, 0 \right) \\
& + \omega_{BOT} \sum_{n=1}^{N} \max_{m=1,2,\dots,8} \left( \sum_{c=1}^{3} t(c, m, n) \right),
\end{aligned} \tag{1}
$$

subject to $t \geq 0$.

Here, the seven terms correspond to target maximum (denoted by TH) and minimum (TL) dose, maximum doses received by the inner (IS) and outer (OS) shells of the target, brainstem maximum dose (BS), mean dose of the ipsilateral cochlea (CO), and total beam-on time (BOT), respectively. $t$ is the beam-on time to be optimized for each sector candidate, that is, each physical sector ($m = 1,2,\dots,8$) with a collimator size ($c = 1, 2, 3$ corresponding to 4-, 8- and 16-mm collimators) at an isocenter candidate ($n = 1,2,\dots,N_{isoc}$). Note that the beam-on time used in Eq.(1) is normalized to a nominal dose rate to ensure the consistent dosimetrical plan quality regardless of GK source decay. $D_i = \sum_j p_{i,j} t_j$ denotes the total dose delivered to voxel $i$, where $p_{i,j}$ denotes the dose contributed by the sector candidate $j$ to voxel $i$ and is precalculated by our in-house developed GK dose calculation engine. $\omega_x$ ($x = TH, TL, IS, OS, BS, CO,$ or $BOT$) denotes the user-specified priority for the corresponding objective, which needs to be adjusted by either a human expert planner or our virtual planner to achieve a desired dose distribution. This objective function can be expressed as a linear programming problem by introducing auxiliary variables [8], and was solved using Gurobi linear programming optimizer (Gurobi optimization LLC., Oregon, US) on CPU in our study. After the optimization phase, sectors with non-zero beam-on time at a same isocenter need to be grouped into composite shots for delivery. In this study, we employed the shot sequencing algorithm developed by Nordström H *et al.* [28], which aimed to maximize the simultaneous radiation delivery in order to minimize the total beam-on time. With this shot sequencing algorithm, the total beam-on time of the composite shots was equal to the total beam-on time of the individual sectors obtained at the optimization phase.

**2.2 DRL-based virtual planner**

In this study, we adopted the DRL scheme that was used by Shen *et al.* for automatic planning of HDR brachytherapy and IMRT [24-27], and developed a virtual planner to automatically adjust those priorities in the plan optimization model during GK inverse planning. Specifically, a priority-tuning policy network was built using deep neural network to parameterize the optimal action-value function for GK inverse planning. The network works in a way analogous to the treatment planning process of human planners. That is, it observes the intermediate GK plan that is obtained given a set of priorities and adjusts the priorities accordingly in an iterative fashion, until a satisfactory plan quality is achieved. More specifically, at each step, the network takes the state of the intermediate plan as input and outputs a priority-tuning action, that is, which priority $\omega_x$ to adjust and how to adjust it. The optimization problem in Eq. (1) with the updated priority set is then solved. The standard RL framework was employed to train the weights of the priority-tuning policy network to approach the optimal action-value function. The details of the priority-tuning policy network and its training process are present in the following subsections.

*2.2.1 Priority-tuning policy network*

To achieve automatic priority tuning, we defined the optimal action-value function in the Q-learning framework [29] as

$$Q^*(s,a) = max_\pi[\sum_{k=0}^{\infty} \gamma^k r_{t+k} \,|s_t = s, a_t = a\,]. \tag{2}$$

Here, $s$ denotes the current state of the intermediate plan generated at a tuning step. In this study, $s$ was a vector composed of the plan quality metrics that were used in our institution for GK plan evaluation, including target coverage, selectivity, dose spillage of 50% prescription dose (referred to as R50), brainstem 0.1cc dose (referred to as $D_{BS,\ 0.1cc}$), cochlea mean dose (referred to as $D_{CO,\ mean}$), and total beam-on time normalized at the nominal dose rate (referred to as $BOT_n$). $a$ denotes the priority-tuning action, that is, which priority $\omega_x$ to adjust and how to adjust. Based on our experiences of manual priority tuning, we have defined 28 tuning action options, as listed in Table 1. $s_t$ denotes the state at the $t_{th}$ tuning step and $a_t$ denotes the action selected at this step. $r_t$ denotes the reward obtained by the action at step $t$. It is positive if the planning objectives are better met, and negative otherwise. $\gamma \in [0,1]$ is the discount factor for the subsequent step. $\pi$ denotes the priority-tuning policy that selects the action which maximizes the value of the $Q^*$ function based on the observed state.

The structure of our priority-tuning policy network, which is a deep neural network, is shown in Figure 1. Specifically, for each action option, a subnetwork was built with two layers of convolutional blocks and three layers of fully connected blocks. Each subnetwork takes the state $s$ as its input, and outputs the value of the $Q^*$ function for the corresponding action. The action whose corresponding subnetwork has the largest output value is the action to be taken at the current tuning step.

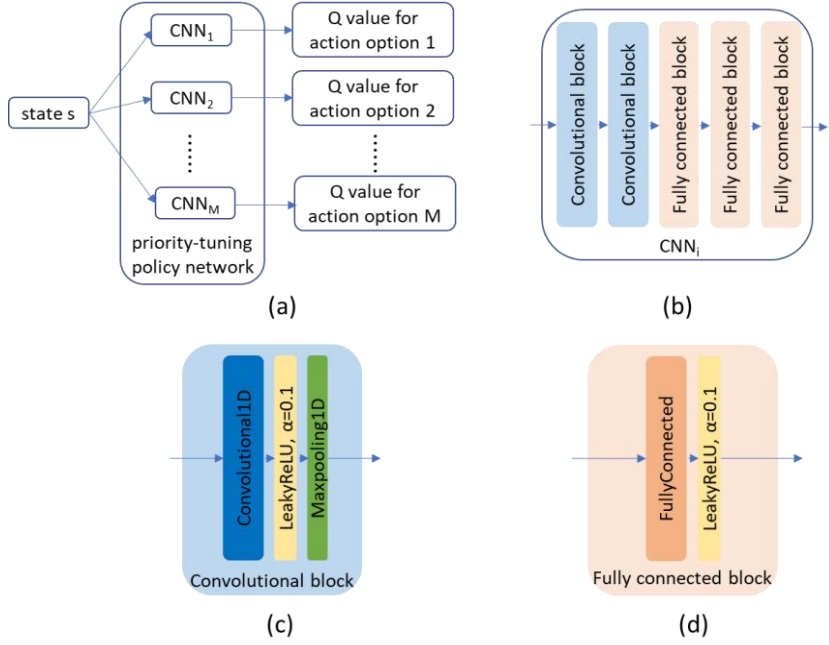

(a)

(b)

(c)

(d)

**Figure 1**. Structure of the priority-tuning policy network. (a) shows the overall network structure, consisting of $M$ Convolutional Neuron Networks (CNN) subnetworks. Each subnetwork corresponds to an action option. (b) shows the subnetwork structure, consisting of two layers of convolutional blocks and three layers of fully connected blocks. (c) and (d) shows the detailed structure of the convolutional block and the fully connected block used in the subnetworks.

Table 1. Tuning action options*

| Action | Action description | Action | Action description |
|---|---|---|---|
| 1 | Increase $\omega_{TL}$ by 10% | 15 | Increase both $\omega_{IS}$ and $\omega_{OS}$ by 10% |
| 2 | Decrease $\omega_{TL}$ by 10% | 16 | Decrease both $\omega_{IS}$ and $\omega_{OS}$ by 10% |
| 3 | Increase $\omega_{TL}$ by 5% | 17 | Increase $\omega_{BS}$ by 50% |
| 4 | Decrease $\omega_{TL}$ by 5% | 18 | Decrease $\omega_{BS}$ by 50% |
| 5 | Increase $\omega_{IS}$ by 50% | 19 | Increase $\omega_{BS}$ by 10% |
| 6 | Decrease $\omega_{IS}$ by 50% | 20 | Decrease $\omega_{BS}$ by 10% |
| 7 | Increase $\omega_{IS}$ by 10% | 21 | Increase $\omega_{CO}$ by 50% |
| 8 | Decrease $\omega_{IS}$ by 10% | 22 | Decrease $\omega_{CO}$ by 50% |
| 9 | Increase $\omega_{OS}$ by 50% | 23 | Increase $\omega_{CO}$ by 10% |
| 10 | Decrease $\omega_{OS}$ by 50% | 24 | Decrease $\omega_{CO}$ by 10% |
| 11 | Increase $\omega_{OS}$ by 10% | 25 | Increase $\omega_{BOT}$ by 50% |
| 12 | Decrease $\omega_{OS}$ by 10% | 26 | Decrease $\omega_{BOT}$ by 50% |
| 13 | Increase both $\omega_{IS}$ and $\omega_{OS}$ by 50% | 27 | Increase $\omega_{BOT}$ by 10% |
| 14 | Decrease both $\omega_{IS}$ and $\omega_{OS}$ by 50% | 28 | Decrease $\omega_{BOT}$ by 10% |

*$\omega_{TH}$ equaled to 1, and was fixed to provide a reference value during tuning.

## 2.2.2 Scoring system

The reward function $r_t$ is to explicitly measure the plan quality change caused by a tuning action. It should be positive if plan quality is improved by the action, and negative otherwise. The planning criteria used in our institution for GK radiosurgery of vestibular schwannoma are: (1) 100% of prescription dose must be received by at least 97% of the target volume, that is, target coverage $\geqslant 0.97$; (2) The maximum dose to 0.1 cc of brainstem must not exceed 12 Gy, that is, $D_{BS,0.1cc} \leqslant 12$ Gy; (3) Try to keep the mean dose of the ipsilateral cochlea below 4 Gy, that is, $D_{co,mean} \leqslant 4$ Gy; (4) Try to obtain selectivity higher than 0.7 and R50 smaller than 5.0 in order to reduce dose spillage to nearby normal tissues, that is, selectivity $\geqslant 0.7$ and R50 $\leqslant 5.0$; (5) Try to reduce total beam-on time for the sake of increasing patient comfort while maintaining acceptable dosimetrical plan quality. At our institution, we usually keep the BOT normalized at the nominal dose rate (i.e., 3 Gy/min) within 60 mins for vestibular schwannoma cases, that is, $BOT_n \lesssim 60$ mins. According to these institutional criteria, we have designed a scoring system $P(s)$ to quantitatively assess the plan quality of the obtained GK plans. The scoring system consisted of a set of scoring functions to evaluate each aforementioned plan metric respectively. The plan score was computed as the summation of the scores for all the plan metrics. The reward function at the $t_{th}$ tuning step can then be calculated as $r = P(s_{t+1}) - P(s_t)$.

Table 2 lists the scoring function for each plan metric, which was designed as a piecewise function composed of two subfunctions, i.e., the arctan() function and a linear function. The arctan() function is for the subdomain where the plan metric of the obtained plan meets our criterion, and the linear function is for the subdomain where the plan metric of the obtained plan doesn't meet the criterion. The plots of these scoring functions are shown in Figure 2. Take the scoring function for target coverage as an example. As shown in Figure 2(a), the arctan() function was used for the target coverages better than 0.97, which was our institutional minimum coverage requirement, and a linear function was used for the target coverages below 0.97. The purpose of using the arctan() function for the target coverages better than 0.97 is to gradually reduce the reward on the improvement of the target coverage when it already meets the criterion and to put more effort on the improvement of other metrics that fall under the criteria. However, when the target coverage is far below 0.97, the arctan() function changes very slightly with the change of the target coverage, which doesn't reward the network much to improve the target coverage. Therefore, instead of the arctan() function, we used a linear function for the target coverage below 0.97 in order to keep rewarding the network to improve the target coverage when it falls under the criterion. With this linear function, the score can be negative if the obtained plan has a very poor target coverage, which alerts the network to the status of target coverage. The scoring functions of other plan metrics have the same format of that of target coverage, with our institutional minimum requirement being used as the cut-off point between the arctan() function and the linear function.

Table 2. The scoring system designed for GK plan quality evaluation

| Plan metric | Scoring function |
| --- | --- |
| target coverage $=$ $\frac{target\ volume\ covered\ by\ prescription\ dose}{target\ volume}$ | score $=\begin{cases} 0.5 + \frac{arctan\left(\frac{coverage-0.97}{0.005}\right)}{\pi}, \ if\ coverage \geq 0.97 \\ 0.5 + \frac{0.25}{0.005} \times (coverage - 0.97), \ if\ coverage < 0.97 \end{cases}$ |
| selectivity $=\frac{target\ volume\ covered\ by\ prescription\ dose}{prescription\ dose\ volume}$ | score $=\begin{cases} 0.5 + \frac{arctan\left(\frac{selectivity-0.7}{0.05}\right)}{\pi}, \ if\ selectivity \geq 0.7 \\ 0.5 + \frac{0.25}{0.05} \times (selectivity - 0.7), \ if\ selectivity < 0.7 \end{cases}$ |
| R50 $=\frac{volume\ of\ 50\%\ prescription\ dose}{target\ volume}$ | score $=\begin{cases} 0.5 + \frac{arctan\left(\frac{5-R50}{0.5}\right)}{\pi}, \ if\ R50 \leq 5 \\ 0.5 + \frac{0.25}{0.5} \times (5 - R50), \ if\ R50 > 5 \end{cases}$ |
| $D_{BS,0.1cc}$ | score $=\begin{cases} 0.5 + \frac{arctan\left(\frac{12-D_{BS,0.1cc}}{0.5}\right)}{\pi}, \ if\ D_{BS,0.1cc} \leq 12\ Gy \\ 0.5 + \frac{0.25}{0.5} \times \left(12 - D_{BS,0.1cc}\right), \ if\ D_{BS,0.1cc} > 12\ Gy \end{cases}$ |
| $D_{CO,mean}$ | score $=\begin{cases} 0.5 + \frac{arctan\left(\frac{4-D_{CO,mean}}{0.5}\right)}{\pi}, \ if\ D_{CO,mean} \leq 4\ Gy \\ 0.5 + \frac{0.25}{0.5} \times \left(4 - D_{CO,mean}\right), \ if\ D_{CO,mean} > 4\ Gy \end{cases}$ |

| $BOT_n$ (in unit of minutes) | $\text{score} = \begin{cases} 0.5 + \frac{\arctan\left(\frac{60-BOT_n}{10}\right)}{\pi}, \text{ if } BOT_n \leq 60 \text{ min} \\ 0.5 + \frac{0.25}{10} \times (60 - BOT_n), \text{ if } BOT_n > 60 \text{ min} \end{cases}$ |
| --- | --- |

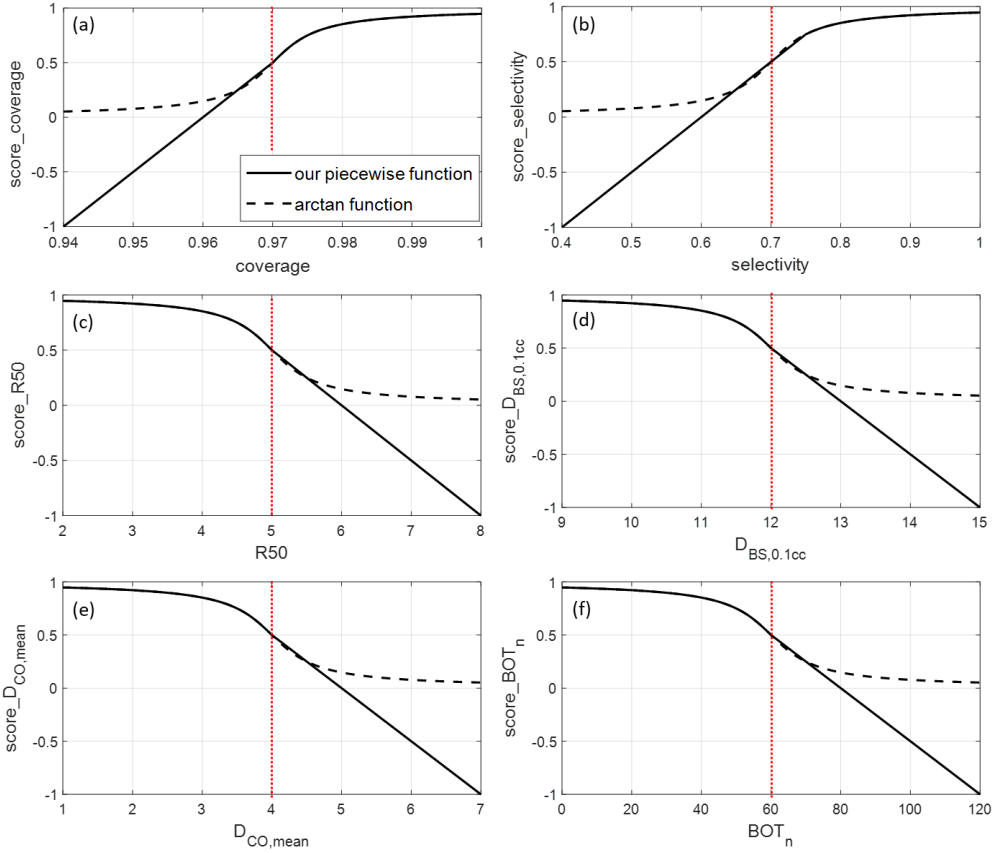

**Figure 2**. Plot of the piecewise functions we used to score each plan metric. The red dotted line denotes the minimum requirement used in our institution. The arctan() function is used for the subdomain where the plan metric of the obtained plan meets our criteria, and a linear function is used for the subdomain where the plan metric of the obtained plan doesn't meet the criteria.

### 2.2.3 Reinforcement learning for network training

The goal of training the priority-tuning policy network is to find a set of network parameters $W$ to satisfy the Bellman equation [30], which is a general property of the optimal action-value function $Q^*(s_t, a_t)$ and formulated as

$$Q^*(s_t, a_t) = r_t + \gamma \, max_{a_{t+1}} Q^*(s_{t+1}, a_{t+1}). \tag{3}$$

This can be achieved by minimizing the deviation of $Q^*(s, a; W)$ from the Bellman equation, which can be formulated as

$$min_W \left[ r_t + \gamma \, max_{a_{t+1}} Q^*(s_{t+1}, a_{t+1}; W) - Q^*(s_t, a_t; W) \right]^2. \tag{4}$$

The workflow of our training process under the standard RL framework is shown in Figure 3. The training process was performed in $N_{epi}$ episodes using $N_{pt\_tr}$ training patient cases. Each episode started with an initial set of the priorities for all training cases and contained a sequence of $N_{step}$ training steps for each case. Specifically, at each training step, the tuning action $a_t$ was determined by the standard $\epsilon$-greedy algorithm with a probability of $\epsilon$, that is, an action was either randomly chosen from all the options with a probability of $\epsilon$ or determined by the currently trained network via $a_t = \text{argmax}_{a_t} Q^*(s_t, a_t; W)$ with a probability of $1 - \epsilon$. Note that the probability $\epsilon$ gradually decayed along the training process, as we gained more and more confidence about the trained network. The optimization problem in Eq.(1) with the updated priority set was then solved to generate a new plan and calculate the new state $s_{t+1}$. The state-action pair $(s_t, a_t, s_{t+1}, r_t)$ obtained at each step was stored into a replay memory to generate a pool of training samples. In order to break the strong correlation between consecutive samples for more efficient learning, we employed the experience replay strategy instead of training our network with the sequential samples. This strategy randomly sampled a small batch of training samples from the replay memory at each step to optimize the network parameters $W$ using Eq. (4). Once finishing $N_{step}$ training steps for the current training case, we moved to the next training case. A new episode started once finishing all the training cases. The whole training process was terminated when the number of episodes reached $N_{epi}$.

During the training, we validated the performance of the trained network every five episodes using $N_{pt\_va}$ validation cases. For each validation case, we started from the initial priority set, performed plan optimization to get the plan state, which was then input into the trained network to determine which tuning action to be taken at this step. The tuning process was terminated after the number of tuning steps reached $N_{step}$. For each validation case, among all the plans generated at each tuning step, the plan with the highest plan score was selected as the final plan. After training, among all the trained networks saved along the training process, the network that achieved the highest mean plan score averaged over all the validation cases were selected as the final model.

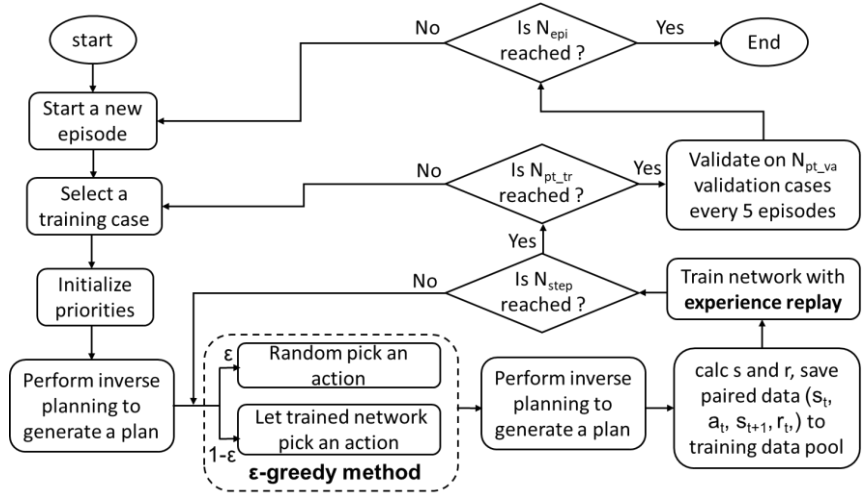

**Figure 3.** Flowchart of our network training process under the standard RL framework.

## 2.3 Evaluations

We have collected data from 26 patients who had vestibular schwannoma and were previously treated at our institution with Gamma Knife radiosurgery. The prescription dose was 12.5 Gy at 50% isodose level. We used 5 cases for training, 5 cases for validation, and the remaining 16 cases for testing. Our network was trained using 150 Episodes with 45 training steps for each case, which took about 7 days on a desktop equipped with an Intel Core 3.3GHz CPU processor, 64GB memory, and an Nvidia GeForce RTX 2080Ti GPU card. With the final model, for each testing case it took less than 2 minutes to finish the inverse planning automatically within 45 tuning steps. The performance of the final model was evaluated by comparing the scores of our final plans with the scores of the initial plans generated using the initial priority set, and the plans obtained by an experienced GK planner via manual priority tuning using the same optimization solver.

## 2.3 Results

The model established at the episode 90 achieved the best performance on the 5 validation cases among all the models saved along the training process, and hence was selected as the final model. The scores of the plans obtained by our final model (referred to as AI plan), the scores of the initial plans obtained with the initial priority set (referred to as initial plan), as well as the scores of the plans obtained via manual priority tuning by an experienced GK planner at our institution (referred to as manual plan), are shown in Table 3-5 for the training dataset, the validation dataset and the testing dataset, respectively. The average scores of the initial plans of these three datasets are $3.63 \pm 1.34$, $3.83 \pm 0.86$, $4.20 \pm 0.78$, respectively. The average scores of the manual plans are $5.28 \pm 0.23$, $4.97 \pm 0.44$, $5.22 \pm 0.26$, and the average scores of the AI plans are $5.42 \pm 0.11$, $5.10 \pm 0.42$, $5.28 \pm 0.20$. These results have shown that our priority-tuning policy network can automatically generate GK plans of comparable plan quality comparing with the plans manually tuned by human expert planners.

**Table 3**. Scores of the obtained plans for the 5 training cases.

| Training cases | Plan score | | |
|---|---|---|---|
| | Initial plan | Manual plan | AI plan |
| R1 | 1.57 | 5.19 | 5.31 |
| R2 | 4.96 | 5.58 | 5.54 |
| R3 | 3.46 | 4.95 | 5.29 |
| R4 | 4.68 | 5.37 | 5.48 |
| R5 | 3.48 | 5.32 | 5.47 |
| Average | $3.63 \pm 1.34$ | $5.28 \pm 0.23$ | $5.42 \pm 0.11$ |

**Table 4**. Scores of the obtained plans for the 5 validation cases.

| Validation cases | Plan score | | |
|---|---|---|---|
| | Initial plan | Manual plan | AI plan |
| V1 | 2.76 | 4.33 | 4.47 |
| V2 | 3.44 | 4.71 | 4.87 |
| V3 | 4.16 | 5.25 | 5.32 |
| V4 | 5.07 | 5.29 | 5.45 |
| V5 | 3.71 | 5.29 | 5.41 |
| Average | $3.83 \pm 0.86$ | $4.97 \pm 0.44$ | $5.10 \pm 0.42$ |

**Table 5**. Scores of the obtained plans for the 16 testing cases.

| Testing cases | Plan score | | |
|---|---|---|---|
| | Initial plan | Manual plan | AI plan |
| T1 | 1.80 | 4.50 | 4.72 |
| T2 | 4.16 | 5.37 | 5.46 |
| T3 | 4.34 | 5.32 | 5.33 |
| T4 | 4.56 | 5.26 | 5.27 |
| T5 | 4.52 | 5.52 | 5.52 |
| T6 | 3.55 | 4.96 | 4.90 |
| T7 | 3.90 | 5.43 | 5.42 |
| T8 | 3.79 | 5.17 | 5.28 |
| T9 | 4.21 | 5.36 | 5.41 |
| T10 | 4.79 | 5.36 | 5.48 |
| T11 | 4.65 | 5.40 | 5.40 |
| T12 | 5.18 | 5.29 | 5.34 |
| T13 | 4.36 | 5.39 | 5.41 |
| T14 | 5.02 | 5.28 | 5.34 |
| T15 | 4.43 | 4.96 | 5.03 |
| T16 | 3.94 | 4.95 | 5.02 |
| Average | $4.20 \pm 0.78$ | $5.22 \pm 0.26$ | $5.28 \pm 0.22$ |

Taking testing case T1 as an example, the evolution of the priorities, plan score, as well as the plan metrics during the tuning process of our final model are shown in Figure 4. The evolution of the dose distribution was shown in Figure 5. It can be observed that the initial plan had a high cochlea mean dose $D_{co,mean}$ and a long normalized total beam-on time $BOT_n$, which didn't meet our criteria. Hence, the priorities $\omega_{CO}$ and $\omega_{BOT}$ were gradually increased by our network to reduce those two metrics until meeting the criteria. Meanwhile, the priorities $\omega_{TL}$ and $\omega_{OS}$ was slightly decreased, which reduced the minimum target dose and the maximum dose of the target's outer shell to slightly compromise on target coverage, selectivity and R50 for the improvement of $D_{co,mean}$ and $BOT_n$. The priority $\omega_{BS}$ was also increased to reduce brainstem dose to further improve the plan score. The plan obtained at the 40th tuning step had the highest plan score and was hence selected as the final AI plan. The values of the plan metrics for the AI plan are shown in Table 6, with those of the initial plan and the manual plan shown in the table as well for comparison. It can be seen that our model reduced $D_{co,mean}$ from 8.7 Gy to 2.0 Gy and $BOT_n$ from 73.7 minutes to 51.8 minutes, while keeping the other plan metrics satisfying the criteria. The manual plan and our AI plan show a very similar dose distribution (as shown in Figure 5 (g) and (h)), and both of them meet all the plan metric criteria (as shown in Table 6).

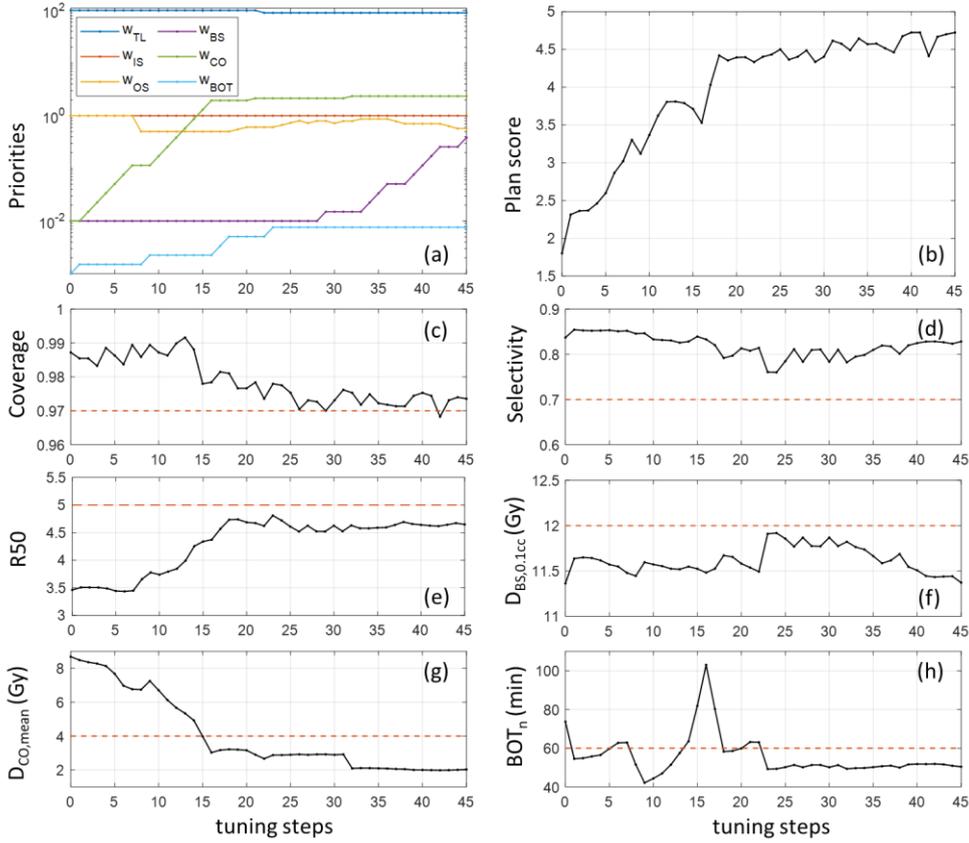

**Figure 4.** Automatic tuning process of the testing case T1 using our final model. (a) shows the evolution of the priorities of the planning objectives during the tuning process; (b) shows the evolution of the plan score obtained by our final model during the tuning process. The plan obtained at the 40th tuning step has the highest score and is hence selected as the final AI plan; (c)-(h) show the evolution of the six plan metrics, i.e., coverage, selectivity, R50, $D_{BS,0.1cc}$, $D_{CO,mean}$, $BOT_n$. The dashed line in (c)-(h) denotes the minimum requirement used at our institution.

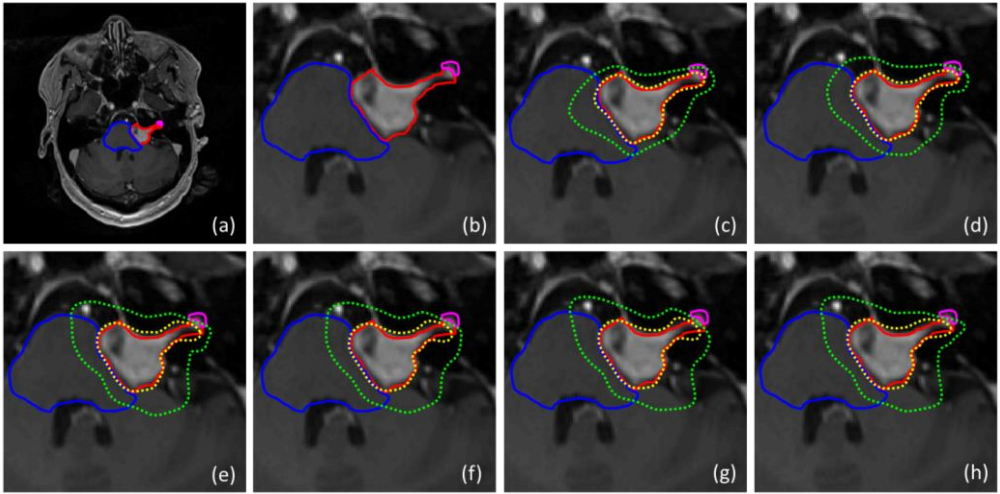

**Figure 5.** Isodose lines of the plans obtained for the testing case T1. (a) shows an axial slice of T1-weighted MR images of the testing case, with the contour of the target shown in blue, the contour of brainstem in blue and the contour of ipsilateral cochlea in magenta; (b) shows the zoomed-in image of the region of interests; (c) shows the isodose lines of the initial plan obtained with the initial priority set (i.e., tuning step = 0). The isodose lines of the prescription dose 12.5 Gy is shown in yellow dotted line, and the isodose line of 50% prescription dose is shown in green dotted line; (d)-(g) show the isodose lines of the plans obtained by our final model at 10th, 20th, 30th, and 40th tuning step, respectively; (h) shows the isodose lines of the plan obtained by our experienced human planner via manual priority tuning.

**Table 6**. Plan metrics of the obtained plans for testing case T1

|             | Coverage | selectivity | R50  | $D_{BS,0.1cc}$ | $D_{CO,mean}$ | $BOT_n$ | Gradient index* |
|-------------|----------|-------------|------|---------|---------|------|-----------------|
| Initial plan | 0.987    | 0.84        | 3.46 | 11.4    | 8.7     | 73.7 | 2.94            |
| Manual plan  | 0.982    | 0.81        | 4.71 | 11.8    | 3.8     | 50.6 | 3.89            |
| AI plan      | 0.975    | 0.82        | 4.64 | 11.6    | 2.0     | 51.8 | 3.90            |

\* Gradient index is an alternative metric to R50 for the evaluation of the dose spillage of 50% prescription dose, and is hence also listed here. It is defined as the ratio of the volume of 50% prescription dose to the volume of 100% prescription dose. It equals to $\frac{R50}{coverage} \times selectivity$.

## 2.4 Conclusions and discussions

In this study, we have built a DRL-based priority-tuning policy network to automatically adjust the priorities among the planning objectives in order to automate GK inverse planning. This study was motivated by the success of applying DRL to automate the plan optimization of HDR brachytherapy and intensity modulated radiotherapy (IMRT) [24-27]. To the best of our knowledge, this is the first study to achieve automatic inverse planning for GK radiosurgery via DRL. Our experimental results demonstrated that our network was able to automatically generate a GK plan comparable or slightly better comparing with the plan generated by human expert planners via manual priority tuning.

Although we employed the linear programming model proposed by Sjölund et al.[8] as the optimization model of GK inverse planning in our study, our DRL-based virtual planner does not depend on the specific optimization model, and can be trained with other inverse planning models. It took about one week to train the network on our desktop workstation under the standard RL framework. A major portion of the computational time was spent on repetitively solving the optimization problem in Eq.(1) with different values of the priorities during the network training. In our study, the optimization problem was solved on the CPU processor of our desktop workstation using Gurobi linear programming solver. A potential solution to accelerate the training process is to develop a GPU-based fast linear programming solver.

In this study, we designed 28 different action options with discrete priority adjustment steps (e.g., 10% and 5% for the priority of target coverage and 50% and 10% for other plan metrics). Theoretically, using a larger number of action options with finer discretization steps would enable finer tuning and might yield higher plan quality. However, it would make the network training more challenging. In our future work, we will attempt to address this issue by employing the deep deterministic policy gradient algorithm in order to take continuous actions[31].

We have designed a set of piecewise scoring functions in this study to evaluate the GK plan quality in order to calculate the reward function, trying to incorporate our institutional GK plan evaluation criteria into the training of our network. These simple scoring functions may not fully reflect the physician's preference and judgement. In order to improve the clinical applicability of our virtual planner, in our future work we will employ the advanced inverse reinforcement learning and imitation learning to first learn the reward function from human expert planners[32,33].

**Acknowledgements**

This study is supported by Winship Cancer Institute #IRG-17-181-06 from the American Cancer Society.